\documentclass[letter]{aa} 
\usepackage{amsmath}	
\usepackage{graphicx}
\usepackage{xcolor}
\usepackage{txfonts}
\usepackage{natbib}
\usepackage{hyperref}
\makeatletter
\renewcommand*\aa@pageof{, page \thepage{} of \pageref*{LastPage}}
\makeatother

\bibpunct{(}{)}{;}{a}{}{,}

\newcommand{\Bdip}{\ensuremath{B_{\rm p}}}

\newcommand{\teff}{\ensuremath{T_{\rm eff}}}
\newcommand{\bz}{\ensuremath{\langle B_z \rangle}}

\newcommand{\bs}{\ensuremath{\langle \vert B \vert \rangle}}

\begin{document}

\title{Metal accretion scars may be common\\ on magnetic, polluted white dwarfs}

\titlerunning{Metal scars on magnetic, polluted white dwarfs}

\author{S.~Bagnulo\inst{1}
\and
J.~D.~Landstreet\inst{1,2}
\and
J.~Farihi\inst{3}
\and
C.~P.~Folsom\inst{4}
\and
M.~A.~Hollands\inst{5}
\and
L.~Fossati\inst{6}
}

\institute{Armagh Observatory \& Planetarium, College Hill, Armagh BT61 9DG, UK\\
\email{stefano.bagnulo@armagh.ac.uk}
\and
Dept. of Physics \& Astronomy, University of Western Ontario, London, Ontario N6A 3K7, Canada
\and
Department of Physics and Astronomy, University College London, London WC1E 6BT, UK
\and
Tartu Observatory, University of Tartu, Observatooriumi 1, T\"oravere, 61602, Estonia
\and
Department of Physics, University of Warwick, Coventry CV4 7AL, UK
\and
Space Research Institute, Austrian Academy of Sciences, A-8042 Graz, Austria
}

   \date{Received June 20, 2024; accepted July 15, 2024}

\abstract{
More than 30\% of white dwarfs exhibit atmospheric metals, which are understood to be from recent or ongoing accretion of circumstellar debris. In cool white dwarfs, surface motions should rapidly homogenise photospheric abundances, and the accreted heavy elements should diffuse inward on a timescale much longer than that for surface mixing. The recent discovery of a metal scar on WD\,0816--310 implies its $B \approx 140$\,kG magnetic field has impeded surface mixing of metals near the visible magnetic pole. Here, we report the discovery of a second magnetic, metal-polluted white dwarf, WD\,2138--332, which exhibits periodic variability in longitudinal field, metal line strength, and broadband photometry. All three variable quantities have the same period, and show remarkable correlations: the published light curves have a brightness minimum exactly when the longitudinal field and line strength have a maximum, and a maximum when the longitudinal field and line strength have a minimum. The simplest interpretation of the line strength variability is that there is an enhanced metal concentration around one pole of the magnetic field; however, the variable line-blanketing cannot account for the observed multi-band light curves. More theoretical work is required to understand the efficiency of horizontal mixing of the accreted metal atoms, and the origin of photometric variability. Because both magnetic, metal-polluted white dwarfs that have been monitored to date show that metal line strengths vary in phase with the longitudinal field, we suggest that metal scars around magnetic poles may be a common feature of metal-polluted white dwarfs.
}

\maketitle

%

\section{Introduction}
Cool white dwarfs are expected to exhibit the presence of only a single element in their spectra, because gravitational diffusion prevails over radiative levitation, and removes all but the lightest element from the atmosphere \citep{fontaine1979,vauclair1979}. In contrast to this theoretical expectation, with sensitive observations, hydrogen- or helium-rich white dwarf atmospheres often show traces of metals \citep{Zucetal03,koester2014}. It is now well established that these photospheric heavy elements originate within rocky planetary bodies \citep{Jura03,farihi2009,klein2010,xu2019,doyle2023} that have survived the post-main-sequence evolution of the host star \citep{veras2016}, and are likely the result of star-grazing orbits that ultimately deliver the material onto the stellar surface \citep[e.g.][]{frewen2014,smallwood2018,malamud2020,akiba2024}. The analysis of the detailed atmospheric abundance pattern of the accreted material provides a unique opportunity to constrain the rocky composition of extrasolar planets and their building blocks \citep[e.g.][]{zuckerman2007,farihi2013,jura2014,melis2017,doyle2019}.

The presence of a magnetic field may have important physical effects on the accretion of planetary material.  Magnetic fields might influence or inhibit the mixing and diffusion of heavy elements in the atmosphere \citep{Cunetal21}, possibly affecting interpretations of chemical compositions and accretion history. A striking example was recently discovered via the metal-rich, magnetic white dwarf (spectral type DZH) WD\,0816--310, which exhibits surface variability of multiple heavy element abundances, as has been deduced from atmospheric modelling. In this star, the concentrated scar of metals has persisted for roughly one diffusion time post-accretion, and is at odds with expectations \citep{Bagetal24}.

Here, we report a second example of a weakly magnetic, polluted white dwarf with a metal scar, WD\,2138--332. (=L\,570-26).  The star also exhibits brightness variations owing to rotation, with a 6.19\,h period and light curve amplitudes in multiple optical bandpasses that are typically of the order of 1\% or less \citep{Faretal24,Heretal24}. The star therefore provides a new opportunity to study relationships between the variable components, but with an additional feature that is lacking in WD\,0816--310.

\section{Observations}
WD\,2138--332 was observed previously using spectropolarimetry to determine the mean longitudinal magnetic field \citep{BagLan18,BagLan19b}.  Having noticed that the metal line strengths had changed between observations taken roughly four years apart, we decided to monitor the star to study the relationship between the magnetic field and line strength variations. New spectropolarimetric observations of WD\,2138--332 were obtained in June and July 2023 with the FORS2 instrument \citep{Appetal98} on the ESO VLT. 

The observations were carried out using the beam-swapping technique \citep[e.g.][]{Bagetal09} using the 1200B grism, which covers the wavelength range from 3700 to 5100\AA. Data reduction was carried out with the FORS2 pipeline \citep{Izzetal10} to obtain two-dimensional wavelength-calibrated frames, from which beams were extracted using standard procedures within {\sc iraf}.

This work also makes use of a spectrum of WD\,2138--332 obtained by us with the ESPaDOnS instrument of the Canada-France-Hawaii Telescope on June 12 2019, and of two ESO archive spectra obtained with FEROS on August 27 and 30 2009. The ESPaDOnS spectrum has $R \simeq 65\,000$, and the FEROS spectra have $R \simeq 48\,000$. 

\section{Measurements}

\subsection{Mean longitudinal magnetic field}\label{Sect_Bz}
\begin{table}
\tabcolsep=0.1cm
\caption{\label{Table_Log} Observing log of FORS2 data. The exposure time was 3000\,s for each
observing series, all consisting of four sub-exposures. }
\begin{center} 
\begin{tabular}{crr@{$\pm$}lr@{$\pm$}l}
\hline\hline
Date                             &            
\multicolumn{1}{c}{UT}           &            
\multicolumn{2}{c}{\bz\ (kG)}    &            
\multicolumn{2}{c}{\bz\ (kG)}    \\           
                                 &            
                                 &            
\multicolumn{2}{c}{(Ca\,{\sc ii})}       &            
\multicolumn{2}{c}{(Ca\,{\sc i}+Fe\,{\sc i})}        \\           
\hline
               2023-06-12 & 05:33 &$ 13.7  $& 1.6  &$ 12.7  $& 0.6  \\%
               2023-06-13 & 05:51 &$ 13.3  $& 1.7  &$ 11.4  $& 0.7  \\%
               2023-06-14 & 05:43 &$ 11.9  $& 1.6  &$  7.4  $& 0.7  \\%
               2023-06-15 & 05:22 &$  5.3  $& 1.6  &$  6.3  $& 0.9  \\%
               2023-06-16 & 05:29 &$ 11.2  $& 1.2  &$  6.1  $& 0.7  \\%
               2023-07-12 & 05:35 &$ 10.7  $& 1.3  &$ 11.9  $& 0.6  \\%
               2023-07-13 & 03:38 &$ 11.8  $& 1.4  &$  6.9  $& 0.7  \\%
               2023-07-13 & 06:50 &$ 12.6  $& 1.1  &$ 10.7  $& 0.6  \\%
               2023-07-18 & 03:30 &$ 12.2  $& 1.8  &$ 13.8  $& 0.6  \\ [2mm]%
               2023-06-29 & 05:29 &\multicolumn{4}{c}{(Linear Polarisation)} \\
               2023-07-23 & 03:28 &\multicolumn{4}{c}{(Linear Polarisation)}\\
 \hline
\end{tabular}
\end{center}
\end{table}

FORS2 circular polarisation spectra were used to measure the mean longitudinal magnetic field, \bz\ , as is explained by \citet{Bagetal24}, and using various metal lines in different wavelength ranges. In Table~\ref{Table_Log}, we report field measurements obtained using the strong Ca\,{\sc ii} H and K lines between 3906 and 4007\,\AA, and those derived from the group of blended metal lines between 4214 and 4415\,\AA, which include a strong Ca\,{\sc i} line and several Fe\,{\sc i} lines. The field measurements made over these two distinct wavelength regions show systematic differences. Keeping in mind that \bz\ is the average longitudinal field, weighted over the visible hemisphere by the local line strengths of each element, the discrepancies between \bz\ values obtained from different ions reflect different surface distributions.  Furthermore, the highly saturated Ca\,{\sc ii} lines respond in a different way to the magnetic field than weaker lines. A better characterisation of the magnetic field may only be achieved with proper modelling of the observed Stokes parameters and simultaneous mapping of the surface distributions of the elements whose lines are used, in a similar way to what is currently done for Ap/Bp stars \citep{DonLan09,Koc20}. In this paper, we use the \bz\ measurements obtained from the Ca\,{\sc i} + Fe\,{\sc i} lines in the 4214 to 4415\,\AA\ range.

\subsection{The mean field modulus}\label{Sect_Bs}
No Zeeman splitting or broadening is detected in the FORS2 intensity spectra, but because of the low spectral resolution, this does not  represent a useful constraint on the mean magnetic field modulus, \bs. From the small line core broadening observed in ESPaDOnS data using H$\alpha$ and the Ca\,{\sc ii} triplet, and in FEROS spectra using H$\alpha$, we deduce that \bs\ is between 40 and 60\,kG (the best fits are $\bs \approx 40$\,kG and 50\,kG to the ESPaDOnS and FEROS data, respectively). 

\subsection{Linear polarisation}
Because linear spectropolarimetry is sensitive to the transverse components of the magnetic field, observations of the linear polarisation of WD\,2138--332 were obtained in two epochs, but no significant signal was detected. \citet{Leoetal17} had suggested that the transverse components of the magnetic field could be probed even from noisy spectra by searching for correlation between $Q/I$ or $U/I$  and the quantity $ 1/I \ {\rm d}^2 I /{\rm d}\lambda^2$, but no correlation could be established from our data. Consequently, our linear polarimetry did not provide a significant constraint on the morphology or the magnetic field.

\subsection{Equivalent width}\label{Sect_EW}
Stokes $I$ (intensity) FORS2 spectra, normalised to the continuum, were used to measure the equivalent width, $W_\lambda$, of spectral lines in various wavelength intervals. Because $W_\lambda$ was measured from all of the individual frames in each observing series, the number of $W_\lambda$ points is $4\times$ that of \bz\ measurements. $W_\lambda$ was also measured from the observations of linear polarisation, but not from the FEROS and ESPaDOnS spectra (due to low S/N). 

\subsection{Atmospheric parameter and abundance determinations}\label{Sect_Abund}
Chemical abundance analysis was carried out on each epoch of FORS2 data of WD\,2138--332, by fitting state-of-the-art atmospheric models \citep{koester2010} to the Stokes~$I$ spectra. In each epoch, spectral models were fitted to the FORS2 spectrum, and all available photometry, considering \teff, $\log g$, radial velocity, and abundances as free parameters.  The combination of the {\em Gaia} parallax and a mass-radius relation \citep{Bedetal20} were used to scale the photospheric models to calculate synthetic photometry at each step. For spectral fits, models were convolved to a resolving power of $R=1400$, and to account for imperfect flux-calibration, each model was normalised against the spectra using a sixth-order polynomial. Optimal parameters were determined using a $\chi^2$-minimization routine, self-consistently recomputing the atmospheric models at each step. 

The adopted stellar parameters are $\teff=7158$\,K, $\log g=8.21$, with chemical abundances [H/He] $-3.79$, [Mg/He] $=-7.12$, [Ca/He] $=-8.23$, and [Fe/He] $=-7.65$ based on the global average.  We note that the inclusion of hydrogen in the model is required as a source of free electrons; it contributes to the He$^-$ free-free opacity, and is needed to fit the observed FORS2 metal line spectra. The synthetic models predict the presence of a weak H$\alpha$ line that is indeed detected in our high-resolution ESPaDOnS spectrum (see Fig.~\ref{Fig_ESP}), confirming a previous detection made by \citet{BagLan21}, based on spectra obtained with the ISIS instrument at the {\it William Herschel} Telescope.  WD\,2138--332 thus has the spectral type DZAH. We note that in the TESS photometric band (from 0.6 to 1.0\,$\mu$m), only the Ca\,{\sc ii} triplet and a weak H$\alpha$ line are present.

\section{Modelling}
\begin{figure}
\begin{center}
\includegraphics[angle=270,width=9.0cm,trim={0.5cm 0.7cm 1.5cm 0.8cm},clip]{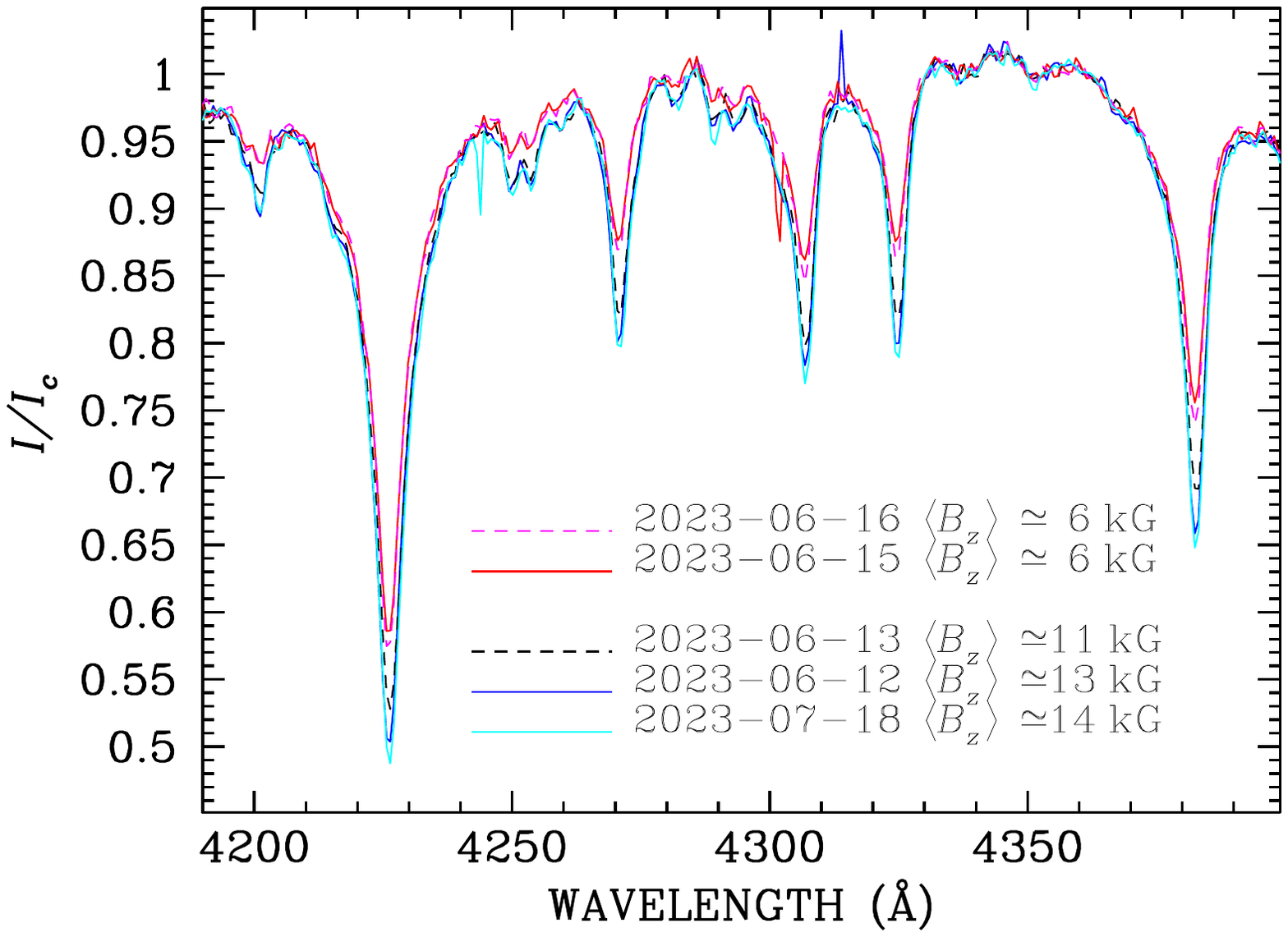}
\end{center}
\caption{\label{Fig_Spectra} Intensity spectra of WD\,2138--332 exhibit metal line variability that is correlated with the mean longitudinal magnetic field, where $W_\lambda$ is stronger when \bz\ is larger, and vice versa.}
\end{figure}
\begin{figure}[ht]
\begin{center}
\includegraphics[angle=0,width=9.0cm,trim={1.4cm 5.1cm 9.6cm 2.9cm},clip]{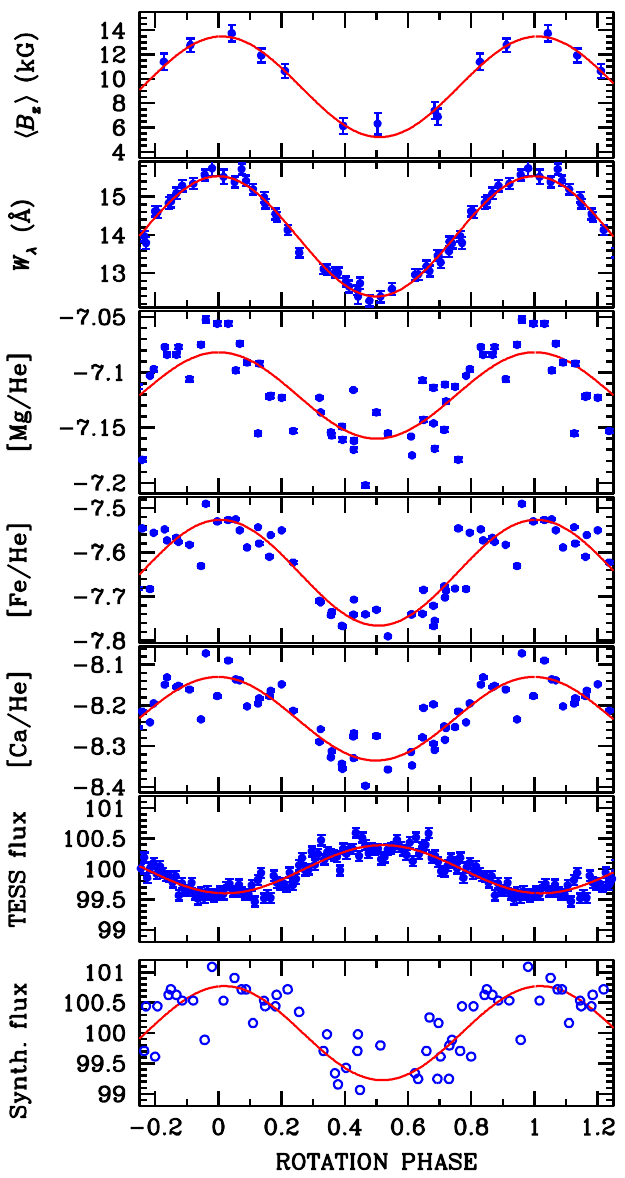}
\end{center}
\caption{\label{Fig_Curves} Multiple metrics for WD\,2138--332 as a function of its rotation phase: the mean longitudinal field, \bz; combined equivalent widths, $W_\lambda$, for the non-saturated metal lines in the ranges $4207-4244$, $4296-4332$, and $4369-4418$; abundance measurements for Mg, Ca, and Fe; and observed and synthetic TESS fluxes (normalised such that their mean values are equal to 100).  In each panel, the best sinusoidal fit is also overplotted.}
\end{figure}

The star shows a variable longitudinal field. Multi-epoch intensity spectra of the star WD\,2138--332 also exhibit clear line variability due to metal abundances that are variable with time (see Fig.~\ref{Fig_Spectra}) . Subtle changes in $W_\lambda$ are detected even within a single observing sequence, from one exposure to the next. 

\subsection{Rotational period}\label{Sect_Period} 
A periodogram analysis was carried out for measurements of the mean longitudinal field, \bz\ , and equivalent width, $W_\lambda$, of the spectral lines in various wavelength intervals. The best ephemeris was obtained from $W_\lambda$:
\begin{equation}
{\rm BJD}_{\rm TDB} = 2460107.7596(10) + 0.25815(3) E
\label{Eq_Ephe}
,\end{equation}
where $T_0$ corresponds to the maximum $W_\lambda$ value in the cycle.

This is consistent with previous period estimates of varying precision derived from space- and ground-based light curves \citep{Faretal24,Heretal24}. Element abundance values were not used to independently attempt to determine the rotational period of the star, but when they are phased with Eq.~(\ref{Eq_Ephe}) they show sinusoidal behaviour with moderate scatter (see Sect.~\ref{Sect_Variations}); no periodic trends are found for \teff, $\log g$, or radial velocity. 

In published light curves based on observations with the Transiting Exoplanet Survey Satellite \citep[TESS;][]{ricker2015}, there is an apparent secondary minimum that can been discerned in phase-folded and binned fluxes, and its partial coverage is obvious in the $u$-band Ultracam light curve \citep[fig.~2;][]{Faretal24}.  This secondary minimum caused a bias in a previously published period analysis, and here this segment was removed from Ultracam $u$-band light curves, which were then combined with TESS and analysed with {\sc period04} \citep{lenz2005}, resulting in a more accurate light curve ephemeris:

\begin{equation}
{\rm BJD}_{\rm TDB} = 2460203.5434(12) + 0.258187(2)E
,\end{equation}
where $T_0$ corresponds to the photometric minimum.  This is consistent with the ephemeris derived from $W_\lambda$, but is an order of magnitude more precise.

\subsection{Magnetic field morphology}
The apparently smooth sinusoidal behaviour of the \bz\ seen in Fig.~\ref{Fig_Curves} demonstrates that the magnetic field must possess a major dipolar component. It has been shown that higher-order field components would contribute only weakly to higher-order terms of the Fourier expansion \citep[e.g.][]{Schwarz1950}; therefore, a non-dipolar component sufficient to account for the observed longitudinal field would certainly produce visible Zeeman splitting. The dipolar component that describes the longitudinal field can completely account for the \bs\ measurements; therefore, if a non-dipolar field component exists, its contribution must be substantially weaker.

The only dipolar models that are consistent with all \bz\ and \bs\ measurements are those with $i \approx 60^\circ$, $\beta \approx 15^\circ$, and $B_{\rm p} \approx 65$\,kG; there is also degeneracy with the solution in which $i$ and $\beta$ are exchanged (see App.~\ref{App_Model}). The uncertainties on the angles are $\approx 5^\circ$, and that of $B_{\rm p}$ is $\approx 5$\,kG.  The conclusion is that the magnetic field at the surface of WD\,2138--332 can be approximated with a relatively weak dipolar field, probably nowhere locally exceeding $\approx 70$\,kG in strength.

\subsection{Phased and anti-phased parameter variations}\label{Sect_Variations}
Figure~\ref{Fig_Curves} demonstrates unambiguous, in-phase correlations between \bz, the $W_\lambda$ of the metal absorption lines, and the abundances of Mg, Fe, and Ca. As in WD\,0816--310, the metal abundances are higher when \bz\ is larger, and vice versa.  Notably, however, all of these parameters vary in anti-phase with the observed TESS light curve, in which the brightness minimum corresponds to the maximum of all other quantities. There is no change in the results if either of the two (spectroscopic and photometric) ephemerides in Sect.~\ref{Sect_Period} are used.

\subsection{Accretion history} 
It is important to consider the accretion history of the star, as it can be compared to the predicted timescale for the horizontal spreading of heavy elements, which should be two to three orders of magnitude shorter than the sinking timescale for WD\,2138--332 \citep{Cunetal21}.  To investigate the composition of the parent material and accretion history, the derived heavy element abundances were combined with appropriate diffusion models for the parameters of WD\,2138--332, including metals and trace hydrogen \citep{koester2020}. Table \ref{Table_Z} provides the calculated element-to-element ratios of the accreted planetary debris for both the increasing and steady-state abundance phases of accretion \citep[see e.g.][]{koester2009}.

There is no indication that Mg is overabundant, relative to undifferentiated chondritic material, suggesting that there has not been enough time for gravitational segregation of elements according to their individual sinking rates during the decreasing abundance phase of accretion \citep{swan2023,Bagetal24}. This is indicated by the first entry in Table~\ref{Table_Z}, in which the elemental ratios for the increasing abundance phase are precisely those currently in the stellar photosphere, and where Ca/Mg is close to chondritic. For both the increasing and steady-state abundance phases, the accreted planetary material appears to be Fe-poor, and possibly modestly enriched in Ca, compared to the bulk Earth and chondrites.  The best match is for mantle-like material in the increasing abundance phase, but where a steady state is modestly favoured in the case of chondritic matter.  Although many polluted white dwarfs accrete material that is most closely matched by chondrites \citep{xu2019,doyle2023}, the additional uncertainty of accretion history (resulting from Myr diffusion timescales) in DZ white dwarfs leads to substantial diversity in observed abundances and interpretations \citep{Hollands2018,swan2023}.

Because the metal sinking timescales for WD\,2138--332 are $\sim 1$\,Myr, a priori there is little chance of observing a $t_{\rm cool}=2.2$\,Gyr white dwarf in the increasing abundance phase of an accretion event (i.e.\ within the first several sinking timescales).  However, in this case it is not possible to robustly distinguish between ongoing, steady-state abundances in which at least several sinking timescales have passed and there has been sufficient time for the metals to horizontally spread, and an increasing abundance phase in which the horizontal mixing has yet to proceed.  
 
\section{Discussion}
WD\,2138--332 is the second weakly magnetic, DZ white dwarf in which accreted metals are found more concentrated around one of the magnetic poles than on the rest of the visible stellar surface. The fact that this feature has been found in the only two magnetic, metal-polluted white dwarfs that have been monitored using spectropolarimetry suggests that the inhomogeneous surface distribution of the heavy elements -- metal scars -- may be common on magnetic, polluted white dwarfs.  

Compared to the prototype WD\,0816--310, the relative amplitudes of the \bz\ and $W_\lambda$ variations are more modest, but this may simply be the result of the geometry of the magnetic field with respect to the line of sight. As the star rotates, WD\,2138--332 is seen by the observer with less contrast than WD\,0816--310. The magnetic poles of WD\,2138--332 never cross the centre of the stellar disc (see App.~\ref{App_Model}), whereas WD\,0816$-$310 may be seen both fully pole-on and equator-on as the star rotates \citep{Bagetal24}. 

The magnetic field of WD\,2138--332 is one of the weakest known to be present at the surface of a white dwarf, with a modulus of $\bs \approx 60$\,kG at most, and provides strong confirmation that even a weak field can influence the observed surface distribution of accreted metals. As in WD\,0816--310, the formation of the metal spots is almost certainly caused by magnetically controlled accretion of ionised gas from a circumstellar disc \citep{Cunetal21,Bagetal24}. However, the persistence of a metal spot at the magnetic pole of a DZ white dwarf is not yet understood. In a non-magnetic DZ star, atmospheric surface convection is expected to distribute accreted atoms uniformly over the stellar surface on a timescale two or more orders of magnitude shorter that the sinking time of the accreted matter \citep{Cunetal21}. Therefore, we should hardly ever observe such a white dwarf with an intact accretion patch. This situation could be modified by a magnetic field sufficiently strong to prevent surface convection entirely; that is, a magnetic field with an energy density higher than the thermal energy density of the local gas \citep{Treetal15,Cunetal21}. However, the required field strength is at least 1\,MG, a lower limit that is more than one order of magnitude larger than the detected field.  In conclusion, it appears that horizontal spreading of heavy elements is affected even when the field should be too weak to impede convective motions at the atmospheric level.  Further theoretical exploration of this situation is essential.

In contrast to the prototype, WD\,2138--332 exhibits detectable, periodic light variations that are synchronised with the magnetic field and spectral line variations. FORS spectropolarimetry and TESS photometry were obtained sufficiently close in time that it was possible to accurately determine the phase relationship of the light curve with \bz\ and $W_\lambda$, and begin exploring causal relations. 

The observed chemical abundance patches, and the model atmospheres computed self-consistently with those abundances, predict that photometric variability should occur. In these models, \teff\ is constant, but the chemical abundances vary, and the variable line (and bound-free) opacity is accounted for. Increasing abundances increases line absorption, particularly in the UV, which increases the effect of line-blanketing and modifies the atmosphere structure. Flux is removed from the UV and redistributed into the visible and infrared. Therefore, in the TESS band, the star should be brighter at rotational phases in which the abundances are larger. This is frequently observed in chemically spotted Ap/Bp stars \citep{Kretal15,Prvetal15,Kretal19}. To quantify this prediction, we calculated synthetic TESS photometry from all our model atmospheres. The amplitude of the predicted variability is of the same order of magnitude as the observed variability; however, the prediction is in anti-phase with the observations. The observed fluxes are faintest when the abundances are largest (Fig.~\ref{Fig_Curves}); the opposite of the prediction. Ultracam observations \citep{Faretal24} in the $ugr$ bands demonstrate similar behaviour to TESS (with a larger amplitude in bluer bands). We conclude that, while variable line opacity and flux redistribution needs to be accounted for in any model fluxes, it cannot explain the observed photometric variability.

The $u$-band light curve obtained with Ultracam shows partial coverage of a secondary minimum starting near the rotational phase 0.35. Hints of a minimum are seen also in the $g$-band and (possibly) in the TESS light curve, but not in $W_\lambda$. We conclude that metal line variations either do not follow exactly the Ultracam $u$-band light curve, or are too small to be firmly detected, while a more dense monitoring of \bz\ will be necessary to establish a detailed correlation (or the lack of it) between the light curve and the longitudinal field.

Brightness variations in magnetic white dwarfs have been known for over three decades \citep[e.g., Feige~7, see][]{Achetal92}. About 20\% of the magnetic white dwarfs in the local 20\,pc volume show light curve variability \citep{Faretal24}, but only WD\,2138--332 has metal spectral lines, suggesting that metal line-blanketing cannot be the direct cause of photometric variability. Investigating the interplay between magnetic field and brightness variations is currently difficult because the phase relationship between the light curve and the magnetic field is almost never established by the available observations.  Besides WD\,2138--332, the best example of a white dwarf for which a secure phase relationship between light curve and magnetic field has been determined is the DA star WD\,1953--011 \citep{Valetal11}, which appears brighter in the $V$ band when the field is weaker, and vice versa (with a phase shift of approximately 0.05 -- 0.1 cycles). In addition, \citet{Lanetal17} presented a \bs\ curve for the DA star WD\,2359--434 that was found to vary approximately in anti-phase with the light curve obtained by \citet{Garetal13} (again with phase shift between 0.05 and 0.1\,cycles).  Compared to WD\,2138--332, both WD\,1953--011 and WD\,2359--434 have magnetic field morphologies that are significantly more complex; in fact, the latter is sufficiently complex that the mean field can only be estimated using the equivalent width of the H$\alpha$ core. Another case in which attempts were made to correlate field morphology and brightness is the DAHe star GD\,356, but the field does not vary sufficiently to establish a clear phase relationship with the light curve \citep{Waletal21}.  

Apart from the case of GD\,356, the observed correlations between the light curve and the magnetic curve may suggest (but not demonstrate) that the magnetic field is responsible for a surface that is frequently photometrically darker where the field is stronger. Indeed, \citet{Valetal14} proposed for WD\,1953--011 a model in which the magnetic field inhibits convection in a spot, creating a region cooler than the rest of the stellar surface. However, atmospheric modelling has shown that the suppression of convection by weak magnetism has no impact on surface temperatures, indicating that rotational modulation of magnetic white dwarf light curves is not due to temperature inhomogeneities.  

Magnetic dichroism could possibly be considered to account for the observed variability of both polarisation and photometry in strongly magnetic, variable stars such as WD\,0912+536 \citep{AngLan71-Periodic,Heretal24,Faretal24}, but it can be hardly responsible for brightness variations  in stars for which the field strength is insufficient to polarise the continuum radiation at a detectable level. Clearly, the interplay between photometric variability, magnetic field morphology, and metal surface inhomogeneities (whenever present) poses significant challenges for theoretical modelling.  

\begin{acknowledgements}
Based on observations obtained with data collected at the Paranal Observatory under program ID 111.24VJ.001 (PI Folsom), and on a spectrum obtained with ESPaDOnS on the Canada-France-Hawaii Telescope (operated by the National Research Council of Canada, the Institut National des Sciences de l’Univers of the Centre National de la Recherche Scientifique of France, and the University of Hawaii), under programme 19AC04 (PI Landstreet). All spectra are available in the relevant observatory archives. JDL acknowledges the financial support of the Natural Sciences and Engineering Research Council of Canada, funding reference number 6377-2016. CPF received funding from the European Union's Horizon Europe research and innovation programme under grant agreement No. 101079231 (EXOHOST), and from the United Kingdom Research and Innovation Horizon Europe Guarantee Scheme (grant number 10051045). This research was partially supported by the Munich Institute for Astro-, Particle and BioPhysics (MIAPbP) which is funded by the Deutsche Forschungsgemeinschaft under Germany Excellence Strategy EXC~2094 – 390783311.
\end{acknowledgements}

\bibliography{sbabib}
\appendix

\section{Detection of \texorpdfstring{H$\alpha$}{Ha}}
\begin{figure}
\begin{center}
\includegraphics[angle=0,width=9.0cm,trim={0.5cm 5.9cm 1.0cm 2.8cm},clip]{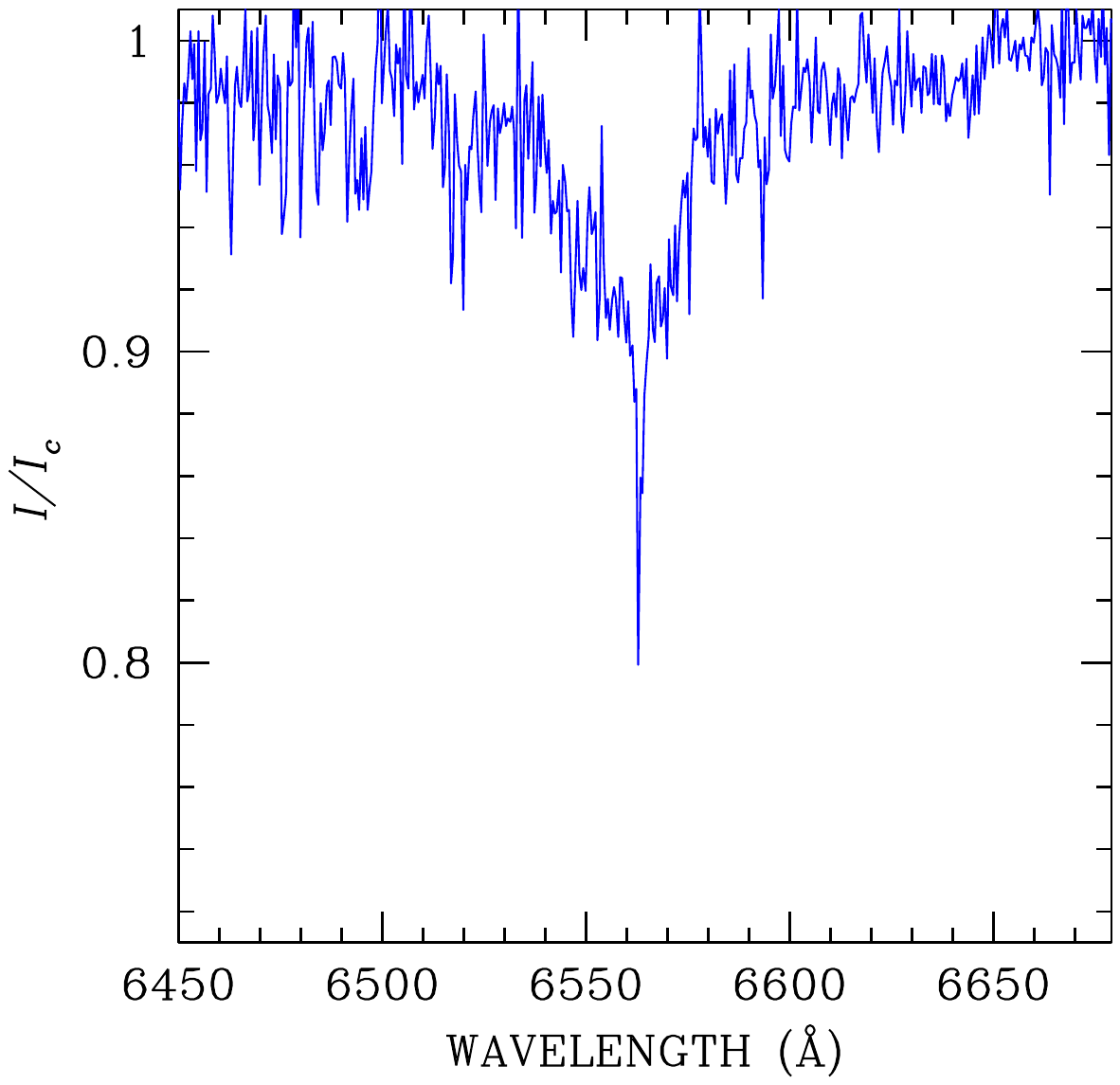}
\end{center}
\caption{\label{Fig_ESP} ESPaDOnS spectrum of WD\,2138--332 in the H$\alpha$ region, smoothed to $R \approx 5000$ to improve the signal-to-noise ratio.}
\end{figure}

Our atmospheric model predicts the presence of a weak H$\alpha$ line, and Figure~\ref{Fig_ESP} shows a clear detection in the ESPaDOnS spectrum. The observed H$\alpha$ has a narrow, non-LTE core that our synthetic models are unable to reproduce, but provides a field estimate $\bs \approx 50$\,kG. 

\section{The magnetic model}\label{App_Model}

Under the assumption of a dipolar field morphology, the \bz\ data are fitted by a function of the rotation phase $f$ that depends on the inclination $i$ of the rotation axis with respect to the line of sight, the angle $\beta$ between the dipole axis and the rotation axis, and the dipolar strength at the magnetic pole $B_{\rm p}$,
\begin{equation}
\bz = 0.31\,B_{\rm p}\ \big[\cos i \cos \beta + \sin i \sin \beta \cos(f-f_0)\big],
\label{Eq_Dip_Bz}
\end{equation}
where $f_0$ is the phase zeropoint. A full analytical description of this model is given, e.g.\ by \citet{Lanetal98}. Under the same hypothesis, the mean field modulus can be approximated by
\begin{equation}
\bs = 0.64 \Bdip\ \left(1+0.25\ \big[ \cos i  \cos \beta + \sin i \sin \beta \cos (f-f_0)\big]^2\right)
\label{Eq_Dip_Bs}
\end{equation}
\citep{Henetal77}.  The numerical coefficients 0.31 in Eq.~(\ref{Eq_Dip_Bz}) and 0.64 in Eq.~(\ref{Eq_Dip_Bs}) are valid for a limb darkening coefficient $u=0.5$ -- analytical expressions valid for any $u$ value are given in \citet{Lanetal98} and \citet{Henetal77}. Data do not offer a strong constraint on the \bs\ value with the rotational phase, but Eq.~(\ref{Eq_Dip_Bs}) shows that \bs\ cannot vary by more than 25\% during a stellar rotation cycle. Therefore we simply search for the best models using the information that the maximum and the minimum \bz\ values are about +14 and +5.5\,kG, respectively, that the maximum of \bs\ is $\ge 45$\,kG, and the \bs\ minimum is $\le 45$\,kG. This of course is an approximate approach, but leads to a quite well-defined result. The best model has ($i$, $\beta$) = ($60 \pm 5^\circ$, $15 \pm 5 ^\circ$) and a dipolar field strength at the pole of $65 \pm 5$\,kG.  Because Eqs.~(\ref{Eq_Dip_Bz}) and (\ref{Eq_Dip_Bs}) are invariant if $i$ and $\beta$ are exchanged, a solution in which the $i$ and $\beta$ values are swapped is also equally acceptable. The uncertainties are estimated by assuming 5\,kG error on the determination of the maximum and minimum \bz\ values. A more sophisticated approach will be possible when more data become available, for example, by obtaining a time series of high-resolution spectra that better constrain \bs. 
 
\begin{figure*}
\begin{center}
\includegraphics[angle=0,width=18cm,trim={3.7cm 6.9cm 1.1cm 12.3cm},clip]{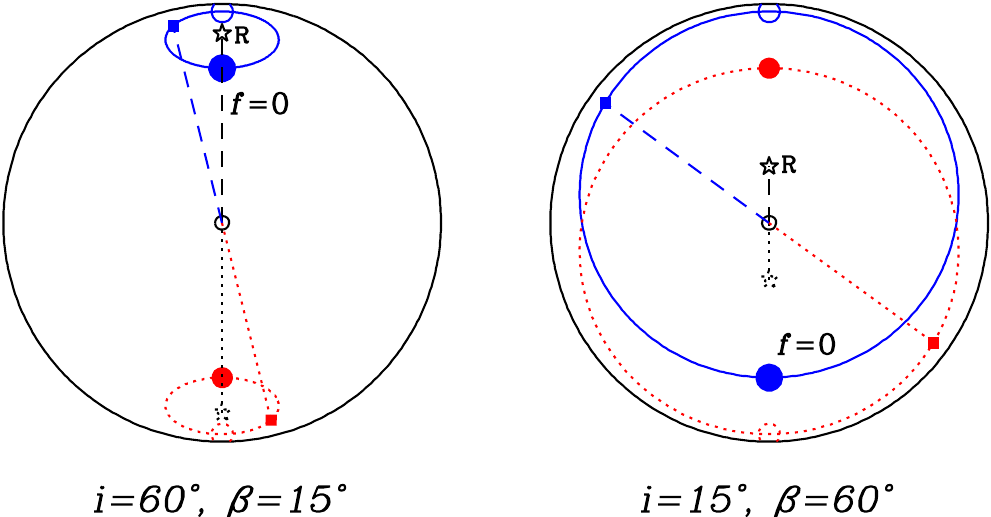}
\end{center}
\caption{\label{Fig_ORM} Representation of the two magnetic models described in the text.
}
\end{figure*}

Figure~\ref{Fig_ORM} shows the position of the magnetic poles on the visible stellar hemisphere as the star rotates, for the two models that cannot be distinguished by the observations. The visible hemisphere is represented by a large black circle. The rotation pole on the visible stellar hemisphere is represented by a black star, marked with the letter “R”; the rotation axis is shown with black line (a dashed line shows the axis crossing the hemisphere closer to the observer, a dotted line represents axis crossing the hemisphere opposite to the observer). The blue solid lines show the trajectory of the north magnetic pole on the visible stellar disc, as the star rotates; the blue solid square shows the position of the pole at an arbitrary phase; in that position, the magnetic axis is shown with a blue dashed line from the centre of the star to the north magnetic pole, and a red dotted line from the centre of the star to the south magnetic pole. The trajectory of the south pole is shown with a red dotted line. The blue solid circle shows the position of the magnetic pole at $f=0$, when \bz\ is maximum, while the empty blue circle shows its position when \bz\ is minimum. The same symbols, in red, refer to the corresponding positions of the south magnetic pole, that is always on the unseen side of the star .

\section{Heavy element mass ratios}
Table~\ref{Table_Z} provides the elemental mass ratios of the parental material as inferred from photospheric abundances and diffusion models, compared with solar system material.

\begin{table}
  \caption{\label{Table_Z}
     Elemental mass ratios for the circumstellar debris accreted by WD\,2138--332.
     The metal ratios for the increasing phase are the current photospheric values,
          as a consequence of the definition of that phase.
    }
\begin{center}
\begin{tabular}{lccc}
\hline\hline

Material                &Ca/Mg &Fe/Mg   &Fe/Ca\\

\hline

Increasing phase        &0.13  &0.68    &5.29 \\
Steady state phase      &0.19  &1.44    &7.54\\

\hline
Chondrites              &0.09  &1.91    &20.1\\
Crust$_\oplus$          &1.64  &1.86    &1.13\\
Mantle$_\oplus$         &0.12  &0.28    &2.42\\

\hline

\end{tabular}
\end{center}
\small {\bf Notes.}  Solar system abundances are from \citet{lodders2003,rudnick2003,palme2003}.
       The mean photospheric abundances (as number rather than mass ratios) are:
         [Mg/He] $=-7.12$, [Ca/He] $=-8.23$, [Fe/He] $=-7.65$.
\end{table}

\end{document}